\documentclass[epjCONF2]{svjour2}
\usepackage{graphics}
\usepackage[varg]{txfonts} 
\usepackage[latin1]{inputenc}
\session-title{}
\begin{document}

\title{Holographic Description of the QCD Phase Diagram And Out of Equilibrium Dynamics
} 
\author{Nick Evans\inst{1}\fnmsep\thanks{\email{evans@soton.ac.uk}}}
\institute{Physics and Astronomy, University of Southampton, Highfield, Southampton, SO17 1BJ, UK.}
\abstract{
Holography provides a powerful tool to model QCD and other strongly coupled gauge dynamics. As examples of its power to compute in novel environments I review some recent results: these include models of the phase structure of QCD at finite temperature and chemical potential; and computation of time dependent processes at strongly coupled phase transitions.\\
{\bf Presented at the International Conference on New Frontiers in  Physics,
ICFP 2012, 10-16 June 2012
Kolymbari, Crete, Greece.}
} 
\maketitle

$\left. \right.$  \vspace{-8.5cm} 

{\hfill 
        SHEP-12-23 \\
} \vspace{7.5cm}

\section{Introduction}

This talk is intended as a small advert for using holography \cite{Maldacena:1997re,Witten:1998qj} to study strongly coupled systems such as QCD. Holography provides new insights into how to model non-abelian gauge theory in environments that were previously beyond our reach. To stress that power I will discuss some examples from my own work \cite{Evans:2010iy,Evans:2011eu,Evans:2012cx,Evans:2010xs}. Firstly I will show that holography allows computation at finite chemical potential and compute the phase diagram of a range of strongly coupled systems. Secondly I shall look at time dependent problems during finite temperature phase transitions. The rigorous examples I shall present are not QCD but other strongly coupled gauge theories (although still with gauge fields and quarks). The ideas behind holography apply equally well to QCD although we do not have a rigorous derivation of the dual - one can nevertheless model build sensibly using the formalism.

\section{Gauge/Gravity Duality}
\label{ggb}

Holography \cite{Maldacena:1997re,Witten:1998qj} provides a unique way to model strongly coupled gauge theories. The renormalization group scale of the gauge theory is treated as a space-time dimension. The conformal symmetry of classical gauge theory is realized as a symmetry of the AdS$_5$ metric
\begin{equation} ds^2 = {r^2 \over R^2} dx_{3+1}^2 + {R^2 \over r^2} dr^2, \end{equation} 
where $R$ is the radius of curvature of the space.
We can think of the space as a box with $r$ corresponding to energy scale. At any fixed $r$ we see
a $3+1$d theory living in the $x_{3+1}$ directions - large $r$ is the UV of theory whilst small $r$ is the IR. The dilatation symmetries in the classical gauge theory act on the space and the fields 
\begin{equation} x \rightarrow e^\alpha x, \hspace{0.5cm} A^\mu \rightarrow e^{-\alpha} A^\mu \end{equation}
but are realized on the AdS space as
\begin{equation} x \rightarrow e^\alpha x, \hspace{0.5cm} r \rightarrow e^{-\alpha} r. \end{equation}
Fields in the AdS space represent gauge invariant operators and sources in the gauge theory such as the gauge coupling $g^2_{YM}$ or $Tr F^2$.

The most clear cut example of this formalism is for the conformal ${\cal N}=4$ supersymmetric gauge theory with a large number of colours (the theory has a gauge field, 6 real scalars and 4 two-component gauginos all in the adjoint representation of SU(N)). In string theory this theory  lives on the world-volume of a stack of D3 branes in the limit where the string tension is taken to infinity. The same D3 branes' tensions source an AdS$_5$ space in the ten dimensional string theory. When there are a large number of D3s ($N \rightarrow \infty$) the radius of the geometry $R \rightarrow \infty$ relative to the string length and a classical gravitational description emerges. Maldacena \cite{Maldacena:1997re} made the remarkable leap of faith that the surface gauge theory and linked closed string theory in AdS$_5$ were alternative descriptions of each other. The full power of this statement is that the weakly coupled 
degrees of freedom in the AdS space (strictly supergravity modes in the tension goes to infinity limit) and their interactions are precisely known from the string theory
and describe the strongly coupled gauge dynamics.

Since the original insight a great deal of work has been done in string theory. It is now believed that the holographic principle works for a wide range of gauge theories \cite{Aharony:1999ti} including all so called deformations \cite{Freedman:1999gp,Freedman:1999gk,Girardello:1998pd} of ${\cal N}=4$ SYM, but also for any theory realizable in the world-volume of branes. In fact it is clear that my first paragraph applies as a starting point for  any gauge theory although there is no assurance that the neccessary deformations of AdS will leave a weakly curved space, nor that one can identify the bulk fields and their interactions. However, holography does provide a new framework for constructing effective descriptions of strong dynamics and there is ample evidence that sensible quantitative results can be found in this way for QCD \cite{Erlich:2005qh,DaRold:2005zs}. Perhaps the most powerful insight is that holography allows computation in environments where other techniques fail. Here we compute at finite density and look at time dependent problems to show the technique's versatility. 

\section{Quarks}

\begin{figure}
\centerline{\resizebox{0.5\columnwidth}{!}{
 \includegraphics{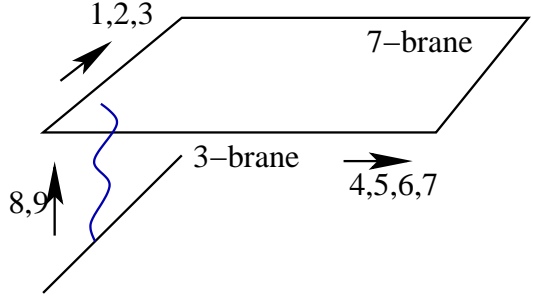} }}
\caption{Quarks can be added to ${\cal N}=4$ SYM as 3-7 strings in the D3-D7 system.}
\label{fig1}       
\end{figure}

Quark fields must be an integral part of any description of QCD. At a phenomenological level we can introduce a scalar field, $\psi$, into the AdS space that is dual to the quark mass and $\langle \bar{q} q \rangle$ condensate. In the UV we can propose the action
\begin{equation} S = \int d^4x ~ dr ~ \sqrt{-g} \left[ (\partial_\mu \psi)^2 - 3 R^2 \psi^2 \right] \end{equation}
which for the AdS background leads to solutions
\begin{equation} \label{psi} \psi \sim {m \over r} + {\langle \bar{q} q \rangle \over r^3}  \end{equation}
the integration constants have the correct dimensions to play the roles of the mass and condensate.
The IR though should be characterized by more subtle interactions of the scalar with itself and in principle the background geometry. 

Top down string constructions \cite{Karch:2002sh,Grana:2001xn,Bertolini:2001qa,Kruczenski:2003be,Babington:2003vm,Erdmenger:2007cm} provide some insight into the form the action could take. Quark ${\cal N}=2$ hypermultiplets can be added to the ${\cal N}=4$ SYM theory through the introduction of D7 branes as shown in Fig \ref{fig1}. The minimum length of the 37 string is a measure of the quark mass (its minimum energy is given by the length times tension). In the quenched $N_f \ll N_c$ limit the D7 branes can be treated as probes in the AdS geometry. The embedding is described by a single function $L(\rho)$ representing the separation of the D3 and D7 in the 89 directions, $L$, as one moves along the D7 in the 
4567 directions, $\rho$. The action for this embedding is given by the Dirac-Born-Infeld action of the brane
(in Einstein frame)
\begin{equation}  \label{DBI}
S_{D7}   ~=~   -T_7 \int d^8\xi e^\phi  \sqrt{- \det P[G]_{ab}} 
 ~=~   -\overline{T_7} \int d^4x~ d \rho ~ \rho^3 e^\phi \sqrt{1 +
(\partial_\rho L)^2}   \end{equation} where $T_7 = 1/(2 \pi)^7
\alpha^{'4}$ and $\overline{T_7} = 2 \pi^2 T_7/ g^2_{uv}$ when we
have integrated over the 3-sphere on the D7. $e^\phi$ is the dilaton that is dual to the gauge coupling. 
Expanding at large $\rho$ one can identify $\psi = L / \rho$. In an AdS background the regular solution (satisfying $\partial_\rho L(0) = 0$) is the flat embedding shown in Fig \ref{fig1}. Massless quarks occur when the D7 lies to
intersect the D3 at $\rho=0$.

\section{Chiral Symmetry Breaking and Confinement}

\begin{figure}
\centerline{\resizebox{0.5\columnwidth}{!}{
 \includegraphics{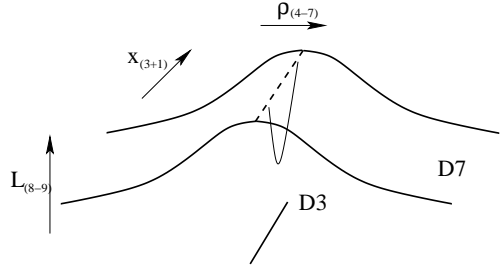} }}
\caption{Schematic of a chiral symmetry breaking D3/D7 construction.}
\label{fig2}       
\end{figure}

An example of a complete top down D3-D7 model that displays chiral symmetry breaking is provided by the imposition of a background magnetic field \cite{Filev:2007gb,Filev:2010pm,Evans:2010iy}. The D7 DBI action contains a world-volume gauge field that is dual to the operators $\bar{q} \gamma^\mu q$  and its source, a background U(1) baryon number gauge field. It is straightforward therefore to study the model in a background B field. In fact the  DBI action for the D7 that results is that in (\ref{DBI}) with an effective un-back-reacted dilaton
\begin{equation} \label{dilB} e^\phi = \sqrt{ 1 + {B^2 \over (\rho^2 + L^2)^2}} \end{equation}
The presence of such a dilaton disfavours D7 embeddings that penetrate to $r= \sqrt{\rho^2 + L^2} = 0$ since the
action grows as $r \rightarrow 0$. The D7 brane therefore bends away from the D3 as shown in Fig \ref{fig2}. 
We can naively see that a configuration that at large $\rho$ (the UV) describes a massless quark is modified at small $\rho$ (the IR) and there in fact is no zero length 37 string. The quark has acquired an effective IR mass due to the chiral symmetry breaking (the asymptotic curvature of the embedding corresponds to the presence of a quark condensate in the spirit of the second solution in (\ref{psi}) ).

Chiral symmetry breaking is induced in QCD by the running gauge coupling which breaks conformal invariance, introduces strong coupling and the scale $\Lambda_{\rm QCD}$. In a phenomenological spirit we can introduce a running coupling by imposing a non-trivial profile for the dilaton (which we will not backreact on the geometry) - for example we can create a step in the dilaton profile \cite{Evans:2011eu}
\begin{equation}
\label{coupling} e^\phi =  A+1 - A \tanh\left[ \Gamma (r - \lambda)
\right] \end{equation}
where $\lambda$ generates the scale equivalent to $\Lambda_{\rm QCD}$, $A$ is the step size, and $\Gamma$ controls the energy regime over which the change occurs. The D7 brane dynamics in this background is similar to the background B field case - the dilaton grows into the IR and repels the D7 brane again triggering chiral symmetry breaking as shown schematically in Fig \ref{fig2}.

The dilaton also enters into the Nambu Goto action for a string. A quark anti-quark pair and their interactions are represented by a 77 string as shown in Fig \ref{fig2} \cite{Rey:1998ik,Maldacena:1998im}. Maldacena \cite{Maldacena:1998im} has shown that in AdS the string droops to deeper values of $r$ the larger the quark anti-quark separation, $d$, is and the string energy scales as  $1/d$ consistent with a Coulomb law in the conformal theory. When a dilaton profile that causes repulsion of the string from the interior of AdS is introduced, the string tends to lie outside the repulsion radius (often called a ``wall''). The energy of the string then grows as $d$ as the quarks are separated since the length of string on the wall is simply lengthened at constant energy cost per unit length. Our phenomenological dilaton profile will therefore lead to confining behaviour too \cite{Evans:2012cx}. 

\section{Temperature and Density}

It has long been known how to deform the AdS space to describe the ${\cal N}=4$ gauge theory at finite temperature \cite{Witten:1998qj}. The correct geometry is an AdS Schwarzschild black hole. The Hawking temperature of the black hole effectively heats up the dual and represents the gauge theory temperature. The geometry is given by
\begin{equation} ds^2 =  {r^2 \over R^2} \left(- f dt^2 + dx_3^2 \right) + {R^2 \over f r^2} dr^2, \hspace{1cm} f= 1 - {r_H^4 \over r^4}  \label{bh} \end{equation} 

When D7 branes are introduced into this geometry a new possible configuration exists in which the D7 brane falls into the black hole horizon \cite{Babington:2003vm,Apreda:2005yz,Mateos:2006nu}. On the gauge theory side this phase represents one where the quark degrees of freedom have become unbound and have merged with the background gluonic thermal bath \cite{Peeters:2006iu,Hoyos:2006gb}. \bigskip

Baryon number chemical potential and density are also easily introduced using the D7 world-volume gauge field previously used to introduce magnetic field \cite{Nakamura:2006xk,Nakamura:2007nx,Kim:2006gp,Kobayashi:2006sb}. One allows a non-zero profile for $A_t$ - for example if it is a constant, $\mu$, then the quark kinetic term replicates a chemical potential term
\begin{equation} \bar{q} \gamma^0 A_t q \rightarrow   \bar{q} \gamma^0 \mu q  \end{equation}
ie a shift in the quark vacuum energy of $\mu$, corresponding to a Fermi surface. 

The relevant gauge field enters the DBI action as
\begin{equation}  \label{DBI2}
S_{D7}   ~=~   -T_7 \int d^8\xi e^\phi  \sqrt{- \det (P[G]_{ab} + F_{ab})} 
\end{equation}
and in the full solutions has UV asymptotic behaviour  $A_t = \mu + {d \over \rho^2}$. Here $d$ represents the quark density operator $\bar{q} \gamma^0 q$. In fact since the action only depends on derivatives of $A_t$ there is a conserved quantity that one may choose, which is precisely $d$. If one sets $d$ then one can solve a single ODE for the embedding $L(\rho)$ and then solve for $\mu(\rho)$ using the $L(\rho)$ solution. This is easier than setting $\mu$ and solving two coupled ODEs for $L(\rho),~A_t(\rho)$.
When the density $d$ is non-zero,   the D7 brane embedding $L(\rho)$ is deformed so that it always ends at $r=0$ \cite{Kobayashi:2006sb}, ie on the D3 branes, or at the black hole horizon if there is also temperature present. The reason is that a non-zero quark density corresponds to strings stretching between the D3 and the D7 - in the infinite tension limit they force the D3 and D7 together.

\section{Phase Diagrams}

We now have all the ingredients we need to determine the phase structure of models of gauge theories with chiral symmetry breaking \cite{Evans:2010iy,Evans:2011eu,Evans:2012cx}. For a given choice of dilaton one can fix $T$ and $\mu$ and then numerically seek regular embeddings of the D7 brane. We will work with massless quarks so that $L \rightarrow 0$ as $\rho \rightarrow \infty$. The configurations that contribute to the phase structure are: the flat embedding, $L=0$, which preserves chiral symmetry; configurations that bend off axis ending off the black hole horizon so that  $\partial_\rho L(0) = 0$, which are chiral symmetry breaking configurations; between these two configurations are solutions that curve off axis but end on the blackhole horizon representing a phase with chiral symmetry breaking and non-zero quark density. Finding all such embeddings across the phase plane is numerical intensive but straightforward. The true vacuum can be found by comparing the configurations' free energy, given by minus the DBI action. 

\begin{figure}
\centerline{\resizebox{0.75\columnwidth}{!}{
 \includegraphics{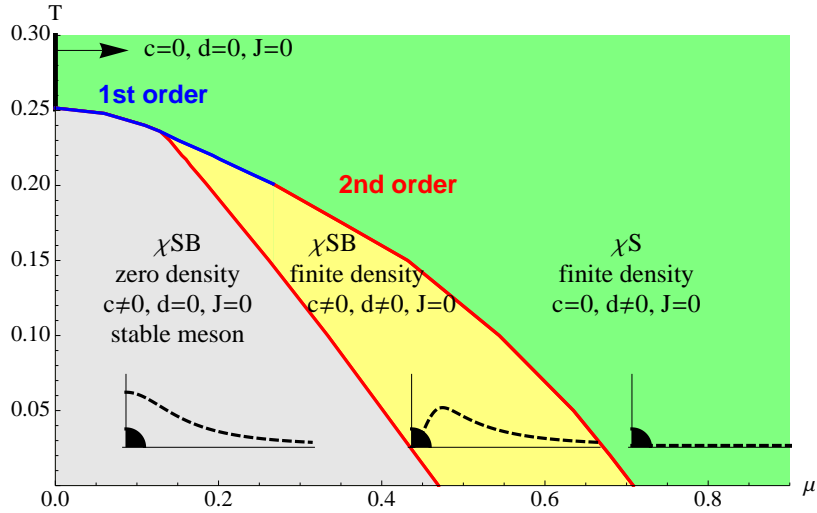} }}
\caption{The phase diagram for the ${\cal N}=4$ SYM theory with ${\cal N}=2$ quark hypermultiplets in the presence of a magnetic field.\cite{Evans:2010iy}}
\label{fig3}       
\end{figure}

In Fig \ref{fig3} we display the phase diagram for the ${\cal N}=4$ SYM theory with ${\cal N}=2$ quark hypermultiplets in the presence of a magnetic field (ie we use the dilaton form in (\ref{dilB}))\cite{Evans:2010iy}. It shows all three phases discussed separated by a complex mixture of first order (blue lines) and second order (red lines)  transitions, including two critical points. It is worth stressing that this is an example of what a powerful technique holography is - no other  method would allow the computation of this phase diagram!

Note that in this framework to move from a chiral symmetry breaking, $d=0$, configuration to the chirally symmetric phase smoothly requires an intermediate phase where the D7 curves off axis and ends on the black hole. That suggests the presence of a phase with a non-zero quark condensate and quark density is very likely in a range of gauge theories.  

I would argue that the B field catalysis of chiral symmetry breaking makes a reasonable first stab at QCD-like dynamics. At these transitions QCD is strongly coupled and characterized by the chiral symmetry breaking scale $\Lambda_{\rm QCD}$ - here we have a strongly coupled gauge theory with a conformal symmetry breaking scale $\sqrt{B}$. Of course our theory has additional degrees of freedom associated with the superpartners but at non-zero T and $\mu$ supersymmetry is broken so there are no special cancellations present in the theory.  The phases and their positions in the plane do indeed match expectations in QCD. The details though are not correct - in QCD one expects a first order transition at low $T$ and high $\mu$ but a second order transition at low $\mu$ and high $T$. 

\begin{figure}
\centerline{\resizebox{1.\columnwidth}{!}{
 \includegraphics{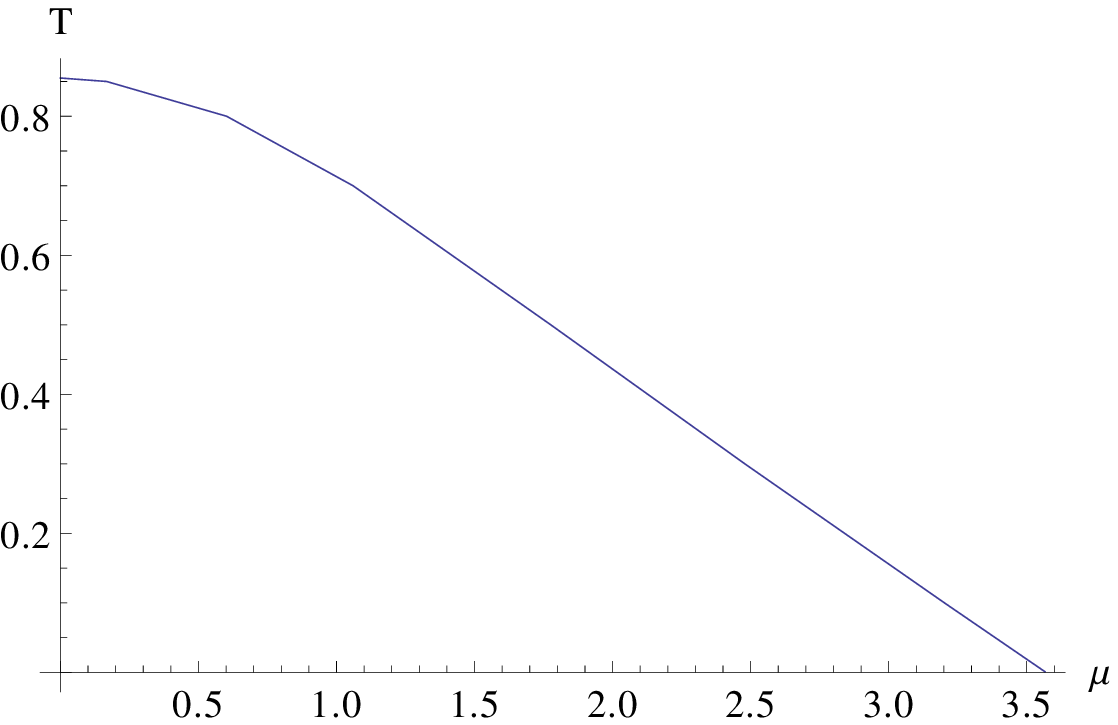} \includegraphics{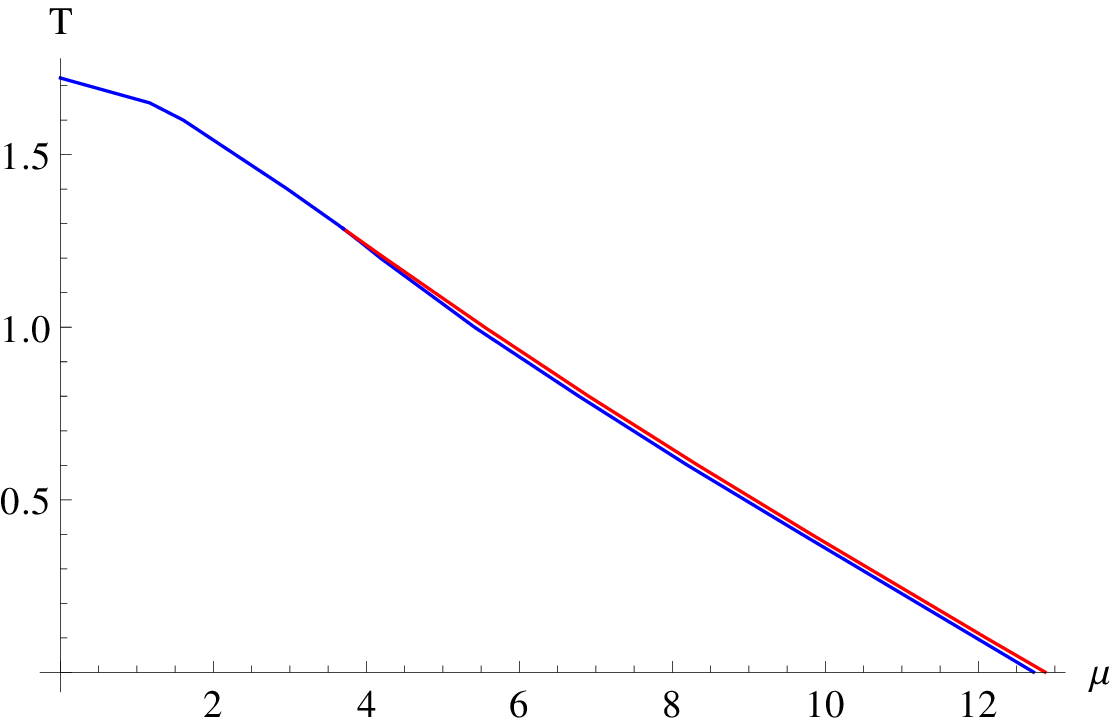}  \includegraphics{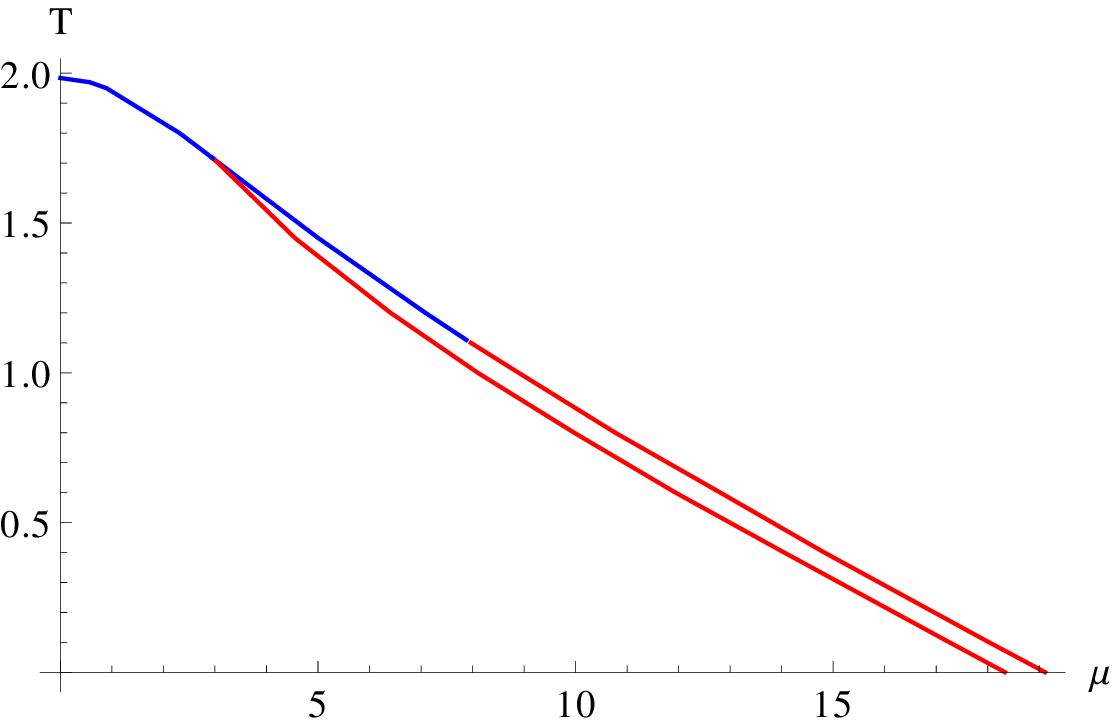}}}
\caption{Plots of three possible phase diagrams for the choices $A=3,15,30$ \cite{Evans:2011eu}.
    Large (small) $A$ gives second (first) order transition
    at low $T$.
    $\Gamma=1, \lambda = 1.715$.}
\label{fig4}       
\end{figure}

It is natural to ask how robust the phase structures of the model are. To test this we will move to the phenomenological ansatz for the coupling in (\ref{coupling}) which allows us to vary the strength of conformal symmetry breaking by hand \cite{Evans:2011eu}. Of course we are now moving away from a precise description to modelling but with many of the key ingredients of the holographic chiral symmetry breaking mechanism intact. The computational methods remain the same \cite{Evans:2011eu}. In Fig \ref{fig4} we show three examples of the phase structure at fixed $\lambda, \Gamma$ but for different values of the step height in the  dilaton ansatz (\ref{coupling}). The three phases correspond in the obvious fashion to those shown in the B field case in Fig \ref{fig3}. The main conclusion is that reducing the strength of the conformal symmetry breaking (ie reducing $A$) in the gauge coupling leads to a preference for a first order chiral phase transition. This model certainly shows that it is relatively easy to engineer a first order transition with chemical potential at zero temperature.

\begin{figure}
\centerline{\resizebox{1.\columnwidth}{!}{
 \includegraphics{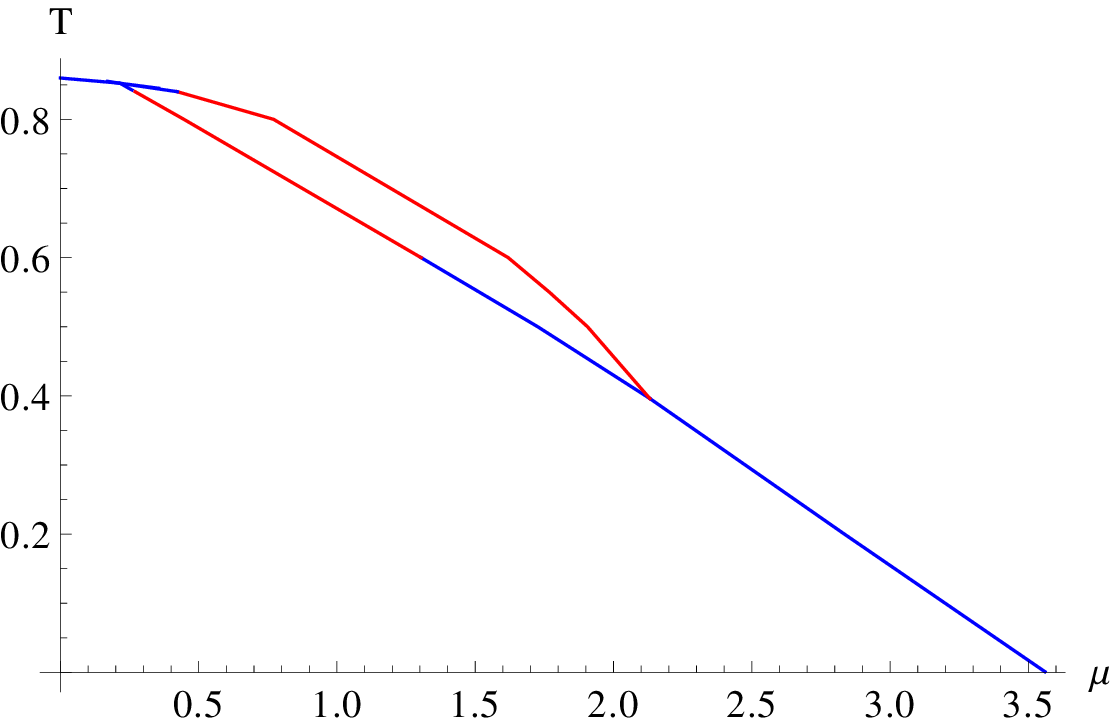}  \includegraphics{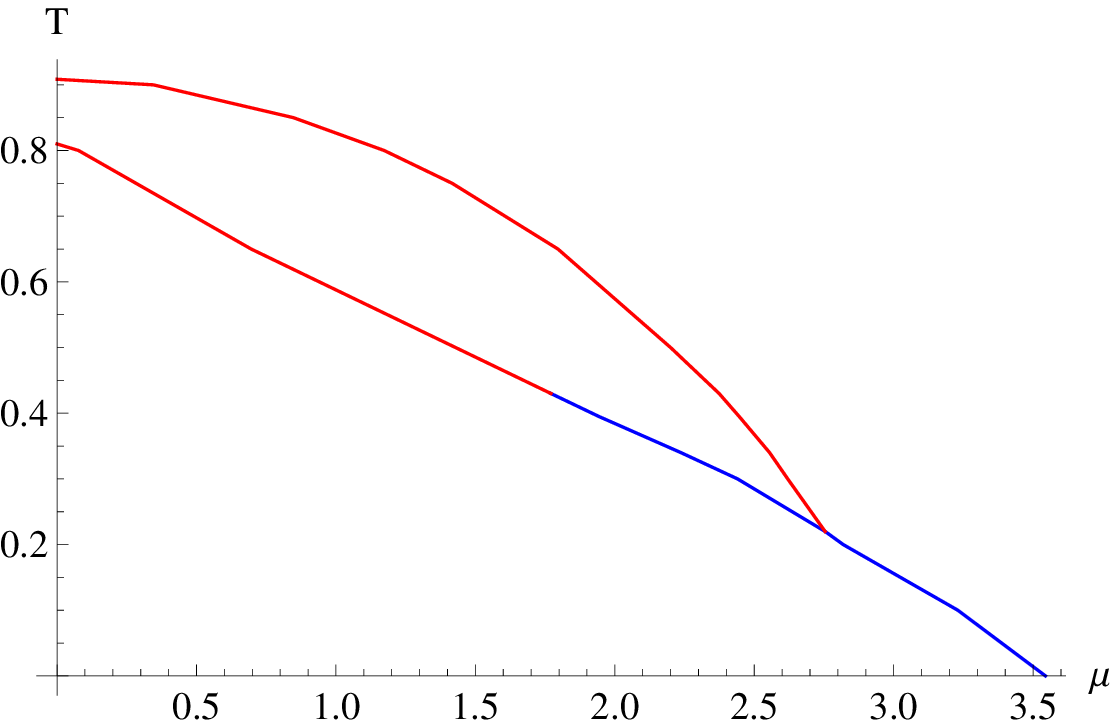}}}
\caption{Sample phase diagrams for theories with none zero $\alpha$ ($\alpha=2.2, \alpha=3$) \cite{Evans:2011eu}.}
\label{fig5}       
\end{figure}

In fact with a little more phenomenological tinkering a second order transition with temperature at low density is also achievable. Away from the probe approximation one should backreact the D7 branes on the AdS geometry. Some serious attempts to include this backreaction for the ${\cal N}=2$ SYM model can be found in \cite{Grana:2001xn,Kirsch:2005uy,Erdmenger:2011bw,Bigazzi:2011db}. Here though in our phenomenological model we simply note that the D7 break the symmetry between the $\rho$ and $L$ directions of the geometry. This symmetry breaking is not present in our black hole ansatz in (\ref{bh}). A simple way to introduce it is to write \cite{Evans:2011eu}
\begin{equation} r^2 =  \rho^2 + L^2 / \alpha \end{equation}
in the emblackening factor $f$ of the black hole geometry. By dialling the phenomenological parameter $\alpha$ one can break the $\rho,L$ symmetry.  If we now repeat our analysis we find the phase diagrams shown in Fig \ref{fig5}. The second order transition at low density in the figure on the right is the main result. This sensibly motivated phenomenological tinkering suggests that the phase structure of strongly coupled gauge theories is quite sensitive to the precise conformal symmetry breaking structure, but equally that it is possible to model a range of phase diagrams using holography. 

\begin{figure}
\centerline{\resizebox{0.5\columnwidth}{!}{
 \includegraphics{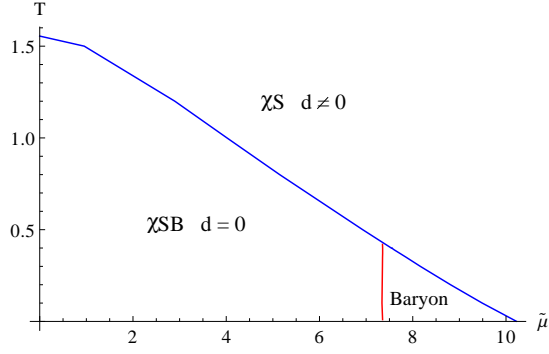} }}
\caption{Sample phase diagram including a baryonic density phase \cite{Evans:2012cx}.}
\label{fig6}       
\end{figure}

A final twist to the story was provided in \cite{Evans:2012cx} where  configurations corresponding to a non-zero density of baryons were
included. Witten described baryons by a D5 brane wrapped on the $S^5$ of the dual geometry \cite{Witten:1998xy,Callan:1999zf}. Such a  configuration must have a surface gauge field sourced by $N$ fundamental strings ending on it to be consistent, suggesting it can be thought of as a baryon. In \cite{Evans:2012cx} new configurations of a D7 ending on such a wrapped D5 (the fundamental strings pull the D5 and D7 to meet at a point, subject to a force balancing condition \cite{Seo:2008qc,Seo:2009kg,Gwak:2012ht}) were included in the model so far discussed. For particular choices of parameter space a QCD-like baryonic phase was achieved as shown in Fig \ref{fig6}. The transition to the phase with baryon density is second order whilst in QCD it is known to be first order at low T - the difference is due to the absence of inter-baryonic interactions in our description (attempts to include such interactions can be found in \cite{Aoki:2012th}).

In conclusion to this section then we see that holography allows the full computation of the phase diagrams of strongly coupled gauge theories. The case of the ${\cal N}=2$ theory with a magnetic field is believed to be a complete computation. Beyond that theory, we have studied how sensitive the phase structure is to phenomenological variation of the conformal symmetry breaking. Broadly the phases present are very robust (and match those expected in QCD). The orders of the transitions are much more sensitive but this does then allow us to cook models of the QCD phase diagram. Further we learn that the weaker the conformal symmetry breaking of a theory the more likely the chiral restoration transition is to be first order. 

\section{Out of Equilibrium Dynamics}

Finally I would like to discuss an additional boon of holographic modelling. It is now relatively easy, using holography, to study time dependent problems in strongly coupled theories since they are mapped to time dependence in weakly coupled classical systems. As
a simple example of the types of computation that are now open to study in \cite{Evans:2010xs} we looked at the thermal, first order, chiral restoration phase transition in the ${\cal N}=2$ gauge theory with a magnetic field. 

We were able to study this system's time evolution thanks to the work in \cite{Janik:2005zt,Nakamura,Janik:2006ft} where a time dependent geometry describing a growing/shrinking black hole in AdS was computed. This
geometry has a black hole whose horizon moves away from the boundary
in time as the ${\cal N}=4$ plasma it describes expands and cools. The time
reversed solution  describes a heating plasma and I will find it
useful to discuss that scenario too below. The geometry is a late
time expansion (when the black hole is small) in powers of inverse
time. By controlling the strength of the $B$ field
on the D7 probes we can arrange to place the
chiral phase transition at any point in the evolution so the
expansion is sufficient to fully study the transition.

The geometry is given by
\begin{equation}
  \frac{ds^2}{R^2} = \frac{1}{z^2} \left( - e^{a(t,z)} dt^2 + e^{b(t,z)} t^2 dy^2
  + e^{c(t,z)} dx_{3+1}^2   \right) + \frac{dz^2}{z^2}  
\end{equation}
where $z=1/r$. At late times, the coefficients can be
expanded to first order  as 
\begin{equation}
\begin{array}{ccc}
  && a(t,z) = \ln \left(\frac{(1-v^4/3)^2}{1+v^4/3}\right)
          +   2 \eta_0 \frac{(9+v^4)v^4}{9-v^8} \left[ \frac{1}{(\epsilon_0^{3/8}t)^{2/3}} \right]
          + \left[\frac{1}{t^{4/3}}\right]  ,    \\
  && b(t,z) = \ln (1 + v^4/3 )
          + \left( - 2 \eta_0 \frac{v^4}{3+v^4} + 2\eta_0 \ln \frac{3-v^4}{3+v^4}\right)
          \left[\frac{1}{(\epsilon_0^{3/8}t)^{2/3}}\right]
          + \left[\frac{1}{t^{4/3}}\right]  ,   \\
  && c(t,z) = \ln (1 + v^4/3 )
          + \left( - 2 \eta_0 \frac{v^4}{3+v^4} - \eta_0 \ln \frac{3-v^4}{3+v^4}\right)
          \left[\frac{1}{(\epsilon_0^{3/8}t)^{2/3}}\right]
          + \left[\frac{1}{t^{4/3}}\right]   , 
\end{array} \end{equation}
with
\begin{equation}
  v \equiv \frac{z}{t^{1/3}} \epsilon_0^{1/4} \ , \qquad \eta_0 = \frac{1}{2^{1/2}3^{3/4}} \ .
  \label{veta}
\end{equation}
$\epsilon_0$ is a free parameter of mass dimension $8/3$ and is related
to the energy density, while $\eta_0$ is related to the shear
viscosity. $v$ is a scaling parameter valid at large $t$. Note
that $a(t,z),b(t,z)$ and $c(t,z)$ are expanded around $t =
\infty$ in powers of $1/t^{2/3}$
as
\begin{equation}
  a(t,z) = \sum_{n=0}^\infty a_n(v) \left( \frac{1}{\epsilon_0^{3/8}t} \right)^{\frac{2}{3}n} \ .\label{abc0}
\end{equation}

Expanding  $g_{tt}$ near the boundary ($z=0$) we can recover a time-dependent emblackening factor, which describes
a moving horizon. The size of the horizon determines the time-dependence
of the `temperature' as~\cite{Nakamura,Grosse}
\begin{equation}
T(t) = \frac{\sqrt{2}}{\pi R^2} r_H = \left( \frac{4\epsilon_0}{3}
\right)^{1/4}\frac{1}{\pi t^{1/3}} \left(
1-\frac{\eta_0}{2\epsilon_0^{1/4}t^{2/3}}   \right) \ . \label{Temp}
\end{equation}

The DBI action for the D7 branes 
schematically reads \cite{Evans:2010xs}
\begin{equation}
  S  \sim  \int d^8\xi \sqrt{- {\rm det} (P[G]_{ab}}) \sim
  \int dt d\rho\,  \rho^3 \mathbb{A}
  \sqrt{
  1+(\partial_\rho L)^2-\mathbb{B} \frac{ (\partial_t L)^2}{(\rho^2 + L^2)^2}  }
  \,,\label{simpleaction}
\end{equation}
with $\mathbb{A}, \mathbb{B}$ complicated but computable functions of $\rho, t$. The evolution of the D7 embedding is now given by a 1+1 dimensional PDE which can be solved by standard numerical packages.

\begin{figure}
\centerline{\resizebox{1. \columnwidth}{!}{
 \includegraphics{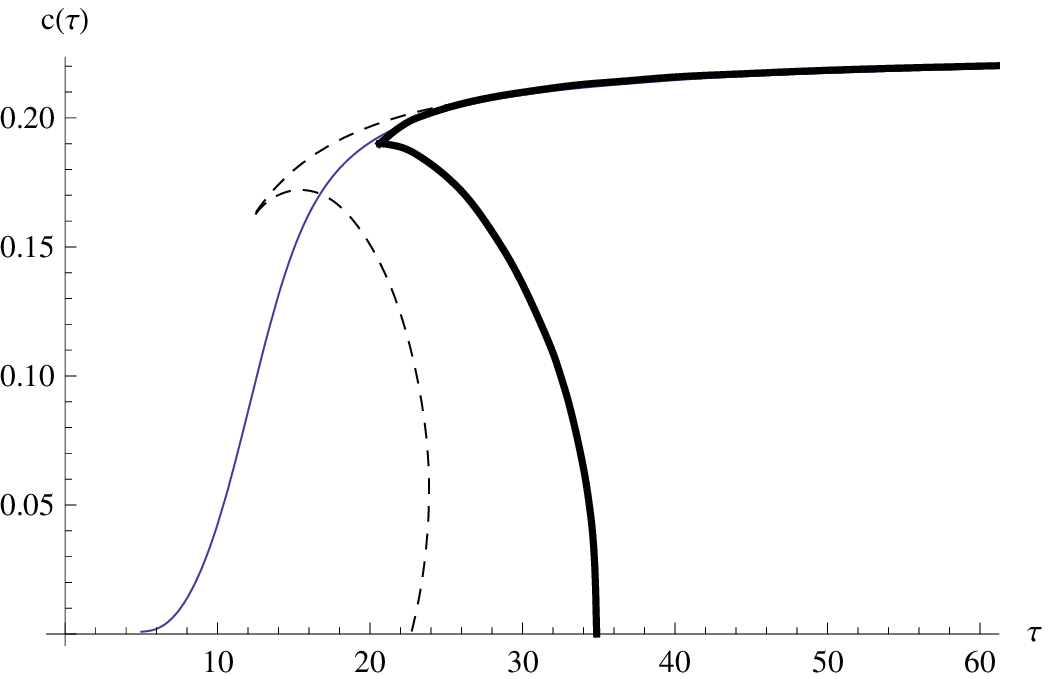} \hspace{1cm} \includegraphics{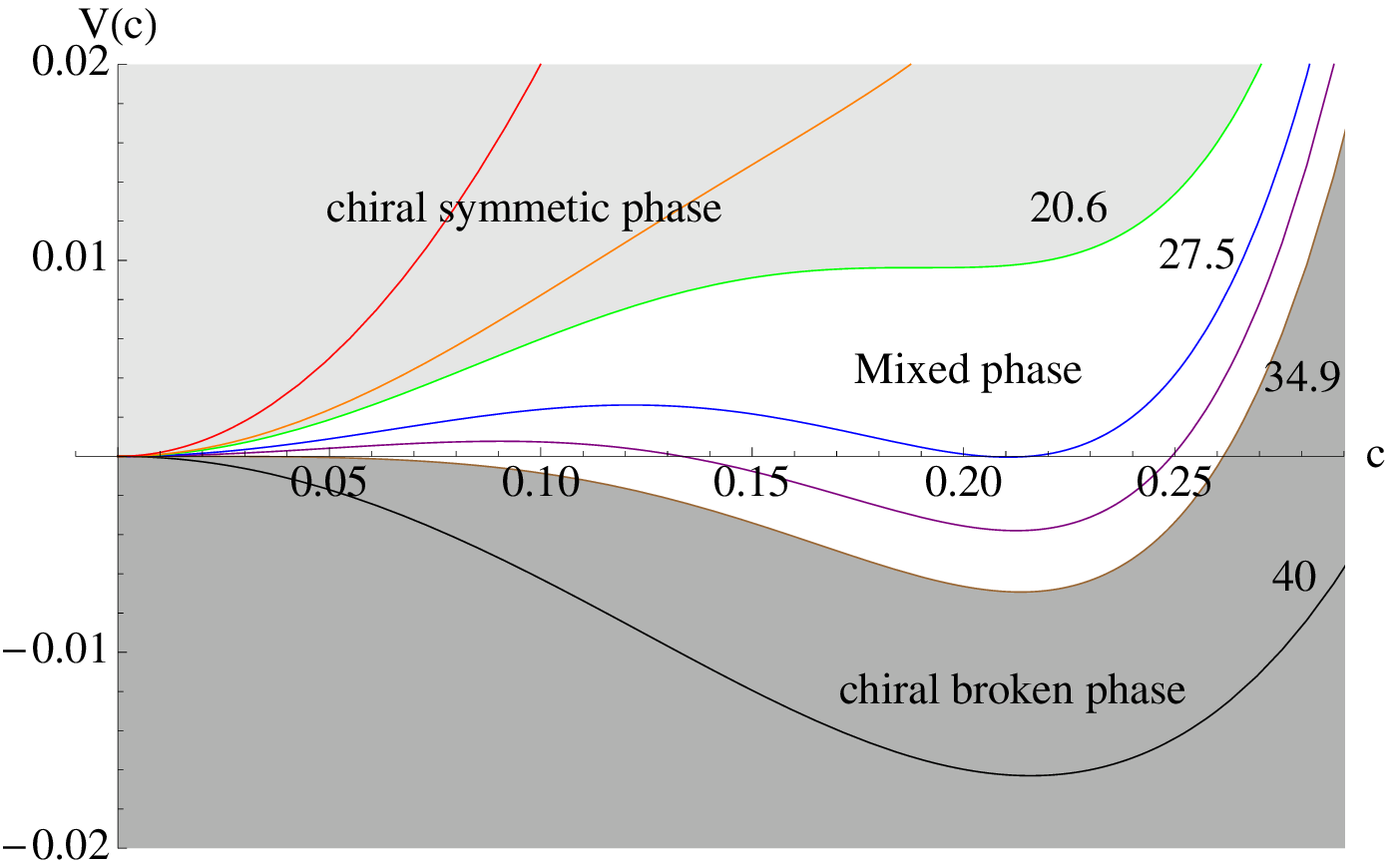} }}
\caption{Results for time dependence of the condensate through the first order phase transition in the ${\cal N}=2$ gauge theory with magnetic field. Left: The condensate $c(t)$ in the equilibrium and non-equilibrium (blue) approaches.
   In the equilibrium description, the different branches correspond to the extrema of the potential $V(c)$ on the Right.  
  \cite{Evans:2010xs}.}
\label{fig7}       
\end{figure}

The simplest example of a quantity one can study is the quark condensate ($c$) as a function of time as the plasma heats or cools. A first approximation is simply to take the expression for $T(t)$ from (\ref{Temp}) and for each time use the equilibrium computation for the condensate (ie use a static black hole of that temperature, find the D7 embedding and read off the condensate from the asymptotic curvature). That computation is represented by the black curve in the left figure of Fig \ref{fig7}. The time axis can be read from left to right for a cooling plasma or right to left for a heating plasma.
At large time the preferred vacuum has a non-zero condensate. Around $t=25$ the first order transition occurs to the chirally symmetric condensate. The behaviour is consistent with the schematic temperature dependence of the potential  shown in the right hand figure in Fig \ref{fig7}. At large times $c=0$ is a maximum of the potential which has a minimum at non-zero $c$. Around the transition there are two minima one with $c=0$ and one with it non-zero, and a maximum between.
At low temperatures the non-zero $c$ minimum disappears and $c=0$ is the unique vacuum. This is a standard first order phase transition with $c$ jumping from a non-zero value to zero around the transition point.

Of course in a real system $c$ does not jump discontinuously. The transition proceeds by bubble formation. The full PDE solutions can be used to study this process. For example if one begins at very large time with the static vacuum configuration at $T=0$ one can watch the evolution as the vacuum is heated. The blue curve in the left hand figure of Fig \ref{fig7} shows the result. It follows the equilibrium expectations until the vacuum at non-zero $c$ ceases to exist at which point it smoothly transitions to the $c=0$ configuration. This represents a super cooled transition. There is no thermal excitation of the vacuum about the minimum at large time, the vacuum sits in the right hand minimum of the potential shown on the right of Fig \ref{fig7} until the temperature where that minimum ceases to exist when it rolls to $c=0$.

\begin{figure}
\centerline{\resizebox{0.5\columnwidth}{!}{
 \includegraphics{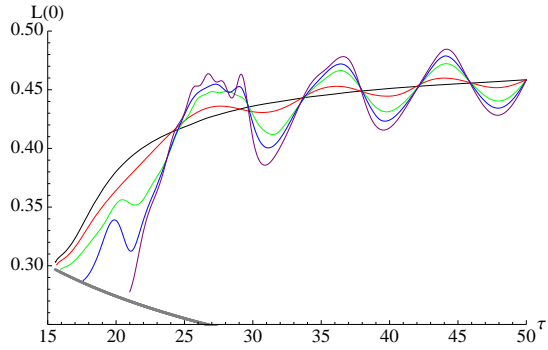} }}
\caption{Plots of the IR position of the D7 brane
against time for a number of large time initial conditions with
different energy \cite{Evans:2010xs} . The thick black line is the horizon.}
\label{fig8}       
\end{figure}

One can add in some thermal excitation by giving the D7 initial condition a small amount of non-zero $\dot{L}$ - at large times it then oscillates about the true vacuum embedding (performs harmonic oscillations about the minimum of the potential at non-zero $c$ in Fig \ref{fig7}). Depending upon how much thermal energy it has, the configuration will move over the hump in the potential near the transition point earlier or later. This is a simplistic description of bubble formation in hotter regions. In Fig \ref{fig8} we plot $L(0)$ for D7 configurations with varying initial thermal energy and we indeed see that they hit the black hole horizon earlier the more energy they have ie they escape to the $c=0$ vacuum earlier in their evolution.

In principle one could study systems with spatially dependent initial conditions for the embedding and one would then observe full bubble formation, bubble collapse before the transition point, and finally bubble growth to drive the transition. These are the  phenomena that are familiar from watching a kettle boil of course, but here, remarkably, we are computing them for a strongly coupled gauge plasma.

\section{Conclusions}

Holography represents a new formalism in which to present renormalization group flow in strongly coupled gauge theories and,
at least for a class of gauge theories derivable from string theory, is known to be a weakly coupled description. It is straightforward to  compute in time dependent problems and in environments with finite chemical potential that are a major obstacle for other approaches. As examples of the technique's power, I have presented models of the phase diagrams for QCD, and more rigorous computations of an ${\cal N}=2$ gauge theory in the presence of a magnetic field, including studies of the first order thermal transition which proceeds via bubble formation.

\newpage

\noindent {\bf Acknowledgements:} I'm grateful for the support of an STFC Rolling Grant and a Durham IPPP Associateship. This work would not have been possible without my collaborators on the various projects: Keun-Young Kim, Astrid Gebauer, 
Maria Magou, Yunseok Seo, Sang-Jin Sin, Tigran Kalaydzhyan and Ingo Kirsch.

\end{document}